**Using Drift Diffusion Model to Analyze Cars' Lane Change Decisions behind Heavy Vehicles**

**Nachuan Li\***
PhD Student
Northwestern University Transportation Center
600 Foster Street
Evanston, IL, 60208, USA
Email: nachuan.li@northwestern.edu

**Hani S. Mahmassani**
William A. Patterson Distinguished Chair in Transportation
Director, Transportation Center
Northwestern University
600 Foster Street
Evanston, IL, 60208, USA
Email: masmah@northwestern.edu

**Soyoung Ahn**
Professor
Department of Civil and Environmental Engineering
University of Wisconsin-Madison
1415 Engineering Drive
Madison, WI, 53706, USA
Email: sue.ahn@wisc.edu

**Anupam Srivastava**
Researcher
Department of Civil and Environmental Engineering
University of Wisconsin-Madison
1415 Engineering Drive
Madison, WI, 53706, USA
Email: asrivastava2@wisc.edu

Word Count: 7138+ 1 table × 250 = 7388 words

Submission Date: Aug 1, 2024

## ABSTRACT

Heavy vehicles (HVs) pose a significant challenge to maintaining a smooth traffic flow on the freeway because they are slower moving and create large blind spots. It is therefore desirable for the followers of HVs to perform lane changes (LCs) to achieve a higher speed and a safer driving environment. Understanding LC behaviors of vehicles behind HVs is important because LCs can lead to highway capacity drop and induce safety risks. In this paper, a drift-diffusion model (DDM) is proposed to model the LC behavior of cars behind HVs. In this drift-diffusion (DD) process, vehicles consider the surrounding traffic environment and accumulate evidence over time. A LC is made if the evidence threshold is exceeded. By obtaining vehicle trajectories with LC intentions in the Third Generation Simulation (TGSIM) dataset through clustering and fitting them with the DDM, we find that a lower initial headway makes the drivers more likely to LC. Furthermore, a larger distance to the follower on the target lane, an increasing target gap size, and a higher speed difference between the target lane and the leading HV increases the rate of evidence accumulation and leads to a LC execution sooner.

**INTRODUCTION**
Due to the high demand of logistic services, it is not uncommon for a car driver to encounter HVs on the freeway. However, the existence of HVs may impede driving safety. Indeed, it has been shown that the percentage of traffic accidents involving trucks is disproportionately high (*1; 2*) and these accidents lead to severe injury (*3-5*). HVs are often moving bottlenecks in traffic streams because they travel slower than cars and have lower acceleration magnitudes (*6-9*). Furthermore, HVs create large blind spots around them due to their large size (*10*). These distinct attributes of HVs lead to cars reacting to them differently compared to when they are around other cars. A stated preference survey conducted by Peeta et al. (*10*) showed that drivers behind HVs prefer to have a larger following distance, reduce their speed, and change lanes if possible. Likewise, Chen et al. (*11*) found through an empirical analysis of the NGSIM dataset (*12*) that even though cars follow trucks with a longer headway, they tend to amplify traffic disturbances. Furthermore, on a 1000ft-long freeway, the number of LCs executed from behind a truck in each hour can be more than 60, which is relatively large.

Since HVs tend to induce different driving behavior and lane changes (LCs) for surrounding cars, it is compelling to understand the LC decision-making process of cars behind HVs. However, modelling LCs behind HVs has been historically challenging because data is generally lacking. Fortunately, the recent introduction of the TGSIM dataset provides a unique opportunity to examine traffic phenomena involving HVs because of its longer duration and higher resolution (*13*). In addition, the drift-diffusion model (DDM), a psychological decision-making model which is based on accumulation of evidence over time (*14*), enables insightful understanding of the processes leading to LC vehicles' decisions. In this study, we develop a DDM framework to model the LC decision-making process of cars behind a HV.

**Background Information**
A close examination into the existing literature indicates that the LC process of vehicles on a freeway has been widely studied. LCs can be classified into Discretionary LCs (DLCs) and Mandatory LCs (MLCs) (*15; 16*), which involve different decision-making processes. While DLCs are often guided by the incentive to gain speed, MLCs are guided by the route choices and obstruction avoidance (*15*). In addition to MLCs and DLCs, Ahmed (*17*) also introduced the forced-merging scheme, which happens when the leader on the adjacent lane slows down to create a gap so that a LC is possible under congested driving conditions.

Rahman et al. (*16*) carried out a comprehensive literature review of microscopic LC models. In all, four categories of microscopic models are identified (*16*): Rule based, discrete-choice based, artificial intelligence based, and incentive based. Rule-based models assume an LC is executed if a set of conditions are met. For example, Gipps (*18*) considered possibility, necessity, and desirability as three key factors of LCs and used a decision tree to model both DLCs and MLCs. Another type of rule-based model is game theory model (*19; 20*), where the merging vehicle chooses whether to perform a LC and the through vehicle chooses whether to give way. Each pair of actions is associated with a different payoff (*19; 20*). Discrete-choice-based models entail the formulation of utility functions and treat LC and no LC as alternatives. For example, in Ahmed (*17*), drivers choose to change lanes with probabilities depending on the traffic conditions and these probabilities are heterogenous across drivers. Given a decision to LC, the probability where a gap is chosen is the probability where both the lead gap and the lag gap exceed the critical gap, which differs across drivers. Toledo, Koutsopoulos and Ben-Akiva (*21*) integrated car-following (CF) and LC behavior into a single model. In their formulation, the target lane and selected gap of the LC vehicles are treated as latent choices. The vehicle first selects the target lane based on its route, each lane's speed and local density. Once the lane choice is made, the vehicle selects a gap based on the gap size and speed difference. Artificial intelligence models are completely data driven and require training with trajectory data. For example, Zhang et al. (*22*) modeled the LC and CF processes of vehicles simultaneously with a long short-term memory model and have shown higher accuracy compared to traditional models. Furthermore, incentive-based models quantify a set of LC desires. For example, the MOBIL model is proposed to minimize the total braking induced by LCs (*23*). First, the safety criterion ensures that the deceleration of a vehicle, which is dependent on the potential gap on the adjacent lane, does not exceed a certain value. If the safety criterion is met, a vehicle would only have the incentive to perform a LC if it is improving the local traffic conditions. In general, polite drivers consider the acceleration of the following vehicles more. In addition to LC models summarized by Rahman et al. (*16*), another category is the duration model. For example, Hamdar and Mahmassani (*24*) introduced a hazard-based model to formulate the duration until the next LC maneuver. In their formulation, the driving behavior is divided into car-following and free-flow episodes, and a LC terminates an episode. Mohammad, Farah, and Zgonnikov (*25*) recently used DDM to model the overtaking time of vehicles on a two-lane highway. The decision of the drivers is made overtime based on the time-to-arrival and location of the oncoming vehicle and the initial speed of the ego vehicle. By fitting their model to a driving simulator dataset, they found that the DDM explains the gap acceptance of the overtaking vehicle relatively well.

While existing studies have provided great insights into the LC decision-making process, few have considered the impact of HVs on the LCs of cars. Toledo, Koutsopoulos and Ben-Akiva (*21*) incorporated the existence of HVs near the LC vehicle as a dummy variable. However, they failed to differentiate the impact of relative locations of the HV on the LC behavior. In fact, when a car is behind a slow-moving HV, it may have a higher desire to perform a LC compared to when it is beside a HV. This implies that a new model is needed for LCs behind HVs. DDM is a widely applied decision-making model in the field of psychology as proposed by Ratcliff (*14*). It treats the decision-making process as evidence accumulation and a decision is made if the evidence threshold for one of the alternatives is reached (*14; 26*). In the field of traffic engineering, DDM has been used to model the pedestrians' decision of when to cross a street with oncoming vehicular traffic (*27-29*) and the gap acceptance of turning vehicles at an unprotected intersection (*30; 31*). However, to the best of our knowledge, the only DDM-based LC model is proposed by Mohammad, Farah, and Zgonnikov (*25*). Unfortunately, their model only contains a few explanatory variables and is not calibrated with naturalistic driving datasets. In addition, their model may not be applicable to LC behavior when overtaking vehicles traveling in the same direction.

Given the limitations of existing studies, our paper aims to address the lacuna in LC models for cars behind HVs. In addition, we use the DDM to model the LC decision-making process, which is relatively new in the field of traffic theory. By training our DDM model to the TGSIM I-395 dataset, we find that a shorter headway to the leading HV, a larger gap to the adjacent-lane follower, an increasing total gap size on the adjacent lane, and a higher speed advantage on the adjacent lane tend to facilitate earlier LC executions. Our study will establish a foundation for more scientific congestion management and accident mitigation strategies on highways with HVs.

**Contribution and Paper Organization**

The contribution of this paper can be summarized as follows:

- We propose a K-means clustering mechanism to sample car trajectories with LC intentions behind HVs
- We develop a DDM to explain the LC decision-making process of cars behind HVs and calibrate it with the TGSIM dataset.

The rest of the paper is organized as follows: "Dataset" section describes the geometry of the data collection site and the process to smoothen vehicle trajectories. "Sample Extraction" section discusses the K-means clustering algorithm to sample trajectories of cars that have an intention to perform a LC from behind the HV. "Drift Diffusion Model" section introduces the DDM framework and elaborates on our model specification. "Result" section presents the DDM parameters and discusses the cars' LC decision-making process. Finally, "Conclusion and Future Research" section summarizes this study and presents future study directions.

## DATASET

In this study, the TGSIM I-395 dataset is used (*13*). It is collected with a stationary helicopter camera in the south of Washington, DC for the duration of approximately 7,000s. Figure 1 shows the freeway geometry, which spans around 468m and contains 6 lanes. Through image processing, each vehicle is tagged with a unique ID and its position in cartesian coordinates is provided every 0.1s. Kalman Filtering is used to smoothen the trajectories (*32*), and the lane boundaries are manually extracted as piecewise-linear lines based on the reference images. The longitudinal locations of the vehicles are obtained through discretization of the freeway into very short segments. This methodology has been discussed by Li et al. (*9*) and Hegde et al. (*33*) in more details. In all, this dataset contains 10,458 unique vehicles that have traveled in total of 97.8 hours. Finally, we only use vehicle trajectories on lane 3, 4, and 5 to reduce the impact of MLCs on our analysis.

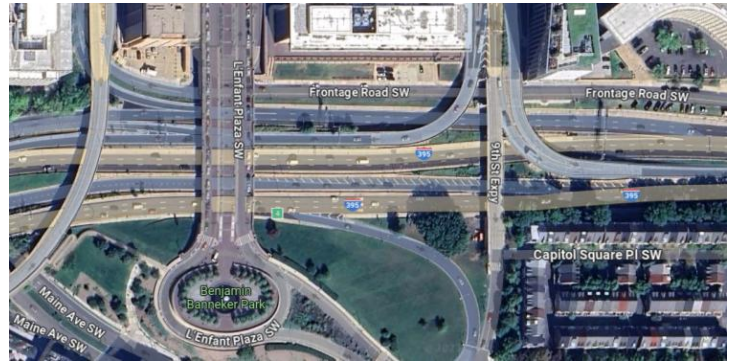

(a)   Satellite Image

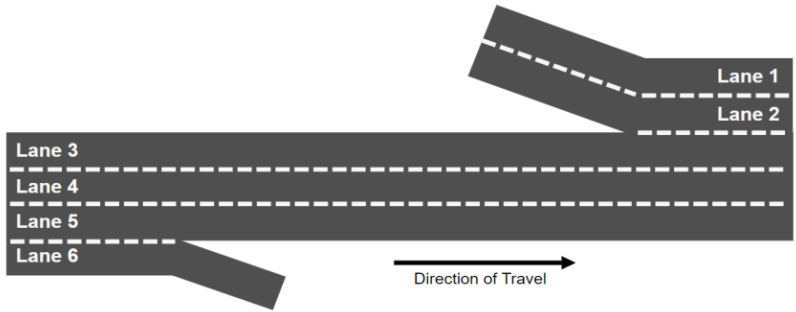

(b)   Road Geometry

**Figure 1** Data Collection Site (*13*)

## SAMPLE EXTRACTION

It is necessary to carefully select the trajectories to be analyzed, because we are only interested in cars that have an incentive to change lanes when they are following HVs. First, we use "central vehicle" to refer to the car that is initially behind the HV. Then, we define a HV-car trajectory pair as the trajectories of the HV and the following car that start and end at the same time steps. The start time of these trajectory pairs are marked by one of the following events:

(1) The central vehicle performs a LC to be behind the HV.
(2) The HV performs a LC to be ahead of the central vehicle.
(3) A car that was between the central vehicle and the HV performs a LC to the adjacent lane.
(4) The preceding trajectory of the central vehicle is upstream of the data collection site and is censored.

On the other hand, one of the following events leads to the end of a HV-car trajectory pair:
(1) The central vehicle performs a LC into the adjacent lane.
(2) The HV performs a LC into the adjacent lane.
(3) A car from the adjacent lane performs a LC to be immediately ahead of the central vehicle.
(4) The succeeding trajectory of the HV is downstream of the data collection site and is censored.

By using the events above to delineate the start and end time, we identified 284 HV-car trajectory pairs with durations of at least 5 seconds, and 41 central vehicles have been observed to perform a LC to the adjacent lane(s). For each central vehicle, we also obtain the location and speed of vehicles on the adjacent lane(s). Figure 2 shows the neighborhood of a central vehicle with two adjacent lanes. Now, we define $d$ as the relative position of a lane to the central vehicle, where $d = 1$ denotes the right adjacent lane and $d = -1$ denotes the left adjacent lane. $G_{L,n,d}$ and $G_{F,n,d}$ are the adjacent lead gap and adjacent follow gap for central vehicle $n$ in lateral direction $d$. $G_{HV,n}$ is the current lead gap between the central vehicle $n$ and the HV.

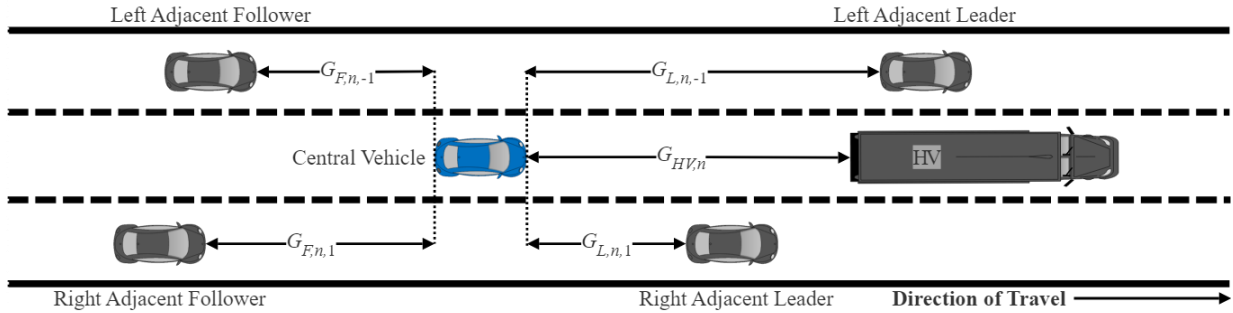

**Figure 2** Local Geometry near the Central Vehicle

It is worth noting that some of the central vehicles may comfortably follow HVs and do not have an intention to perform LCs. Including central vehicles with no LC intention into DDM fitting may lead to erroneous parameter estimation since they are not deciding when to LC. Therefore, it is important to exclude them. Obtaining the LC intention of a central vehicle is challenging because it cannot be directly observed based on the vehicle trajectory data. However, central vehicles that have a desire to change lanes may have different behaviors compared to those that prefer to follow the HVs. Therefore, we extract relevant features from the HV-car trajectory pairs and group them into two clusters. If most trajectory pairs where the central vehicle has performed a LC fall in the same cluster, then it is likely that central vehicles in this cluster have LC intention.

In all, five longitudinal features are extracted from each trajectory pair:
- The duration, denoted as $T_n$
- The average gap between the central vehicle and the HV, denoted as $\overline{G_{HV,n}(t)}$
- The standard deviation of the current lead gap size over time, denoted as $\sigma_{G_{HV,n}}(t)$
- The average acceleration of the central vehicle, denotes as $\overline{a_n(t)}$
- The average acceleration of the HV, denoted as $\overline{a_{HV,n}(t)}$

$$D_n = \overline{G_{HV,n}(t)} + \gamma_1 \frac{\sigma_{G_{HV,n}}(t)}{T_n} + \gamma_2 \left( \overline{a_n(t)} - \overline{a_{HV,n}(t)} \right) \qquad \text{Eq.1}$$

We use Eq.1 to map the five features into a single feature position $D_n$. Instead of $\sigma_{G_{HV,n}}(t)$, $\frac{\sigma_{G_{HV,n}}(t)}{T_n}$ is used because the standard deviation of the current lead gap is positively correlated with the duration of the trajectory pair, which makes a normalization necessary. $\overline{a_n(t)} - \overline{a_{HV,n}(t)}$ is the difference between the average acceleration of the central vehicle and the HV. $\gamma_1$ and $\gamma_2$ are weight parameters to be obtained.

Features associated with the adjacent lanes are not used because they are included in the DDM and using them in clustering may lead to self-selection bias for DDM parameter estimation.

Finally, K-means clustering is used to assign the trajectory pairs into 2 clusters based on the feature positions. We find $\gamma_1$ and $\gamma_2$ that minimize the sum of squared distance from each feature position to its cluster center. By using the BFGS optimizer in SciPy (*34*), we obtained $\gamma_1 = -23.62$ and $\gamma_1 = -4.27$, which are both significant under the 95% confidence level.

Figure 3 shows the distribution of the feature positions for the two clusters. In all, cluster 1 has a mean feature position of 12.506. It contains 265 HV-car trajectory pairs, and all 41 central vehicles that have performed a LC are included. Cluster 2 has a mean feature position of 102.540 and contains only 16 central vehicles that have not performed a LC. Since all LC vehicles fall in cluster 1, the non-LC central vehicles in it may have a similar longitudinal driving behavior to them. This may imply that vehicles in cluster 1 likely have LC intention, even though most of them have not executed a LC in the observation period. To better fit our DDM, we only use central vehicles in cluster 1, which likely excludes vehicles with no LC intention.

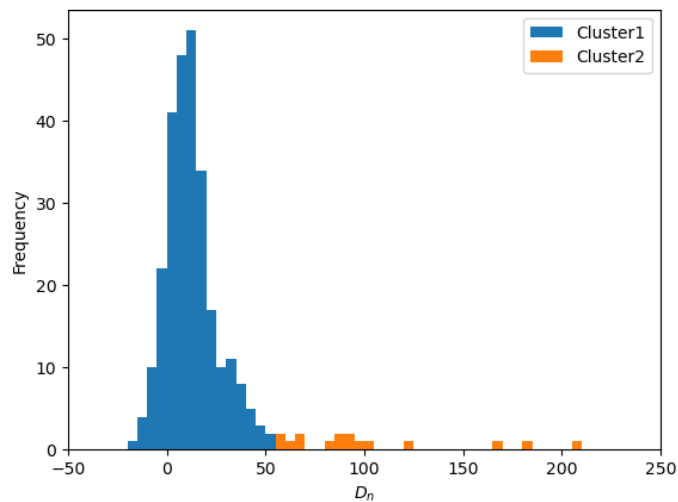

**Figure 3** Distributions of Clustering Metrics

Based on Eq.1 and the values of $\gamma_1$ and $\gamma_2$, central vehicles in cluster 1 tend to have smaller but more time-variant gap size to the HV. In addition, the difference between the average acceleration of the car and that of the HV seems larger. This may be because gaining speed is a key incentive for the central vehicle to perform a LC: when following a slower HV, they may tailgate the HV and look for a gap on the adjacent lane by adjusting their speed. However, they may have less incentive to keep the same current lead gap size or the same speed as the HV. For example, the trajectory pairs in Figure 4(a) and Figure 4(b) belong to cluster 1. In both cases, the gap between the central vehicle and the HV is relatively small. In Figure 4(a), the central vehicle might have picked a gap slightly downstream and eventually managed to merge into it by decelerating. In Figure 4(b), while no LC is eventually made, the central vehicle follows the HV very closely with a decreasing gap size. This may imply impatient driving behavior and a desire to overtake. Figure 4(c) and Figure 4(d) show trajectory pairs in cluster 2. In Figure 4(c), the central vehicle is far from the HV, where a LC to improve speed is likely not needed even though the HV is slower. In Figure 4(d), both the central vehicle and the HV slightly decelerate, and the gap size remains relatively constant, possibly implying that the central vehicle is comfortably following the HV.

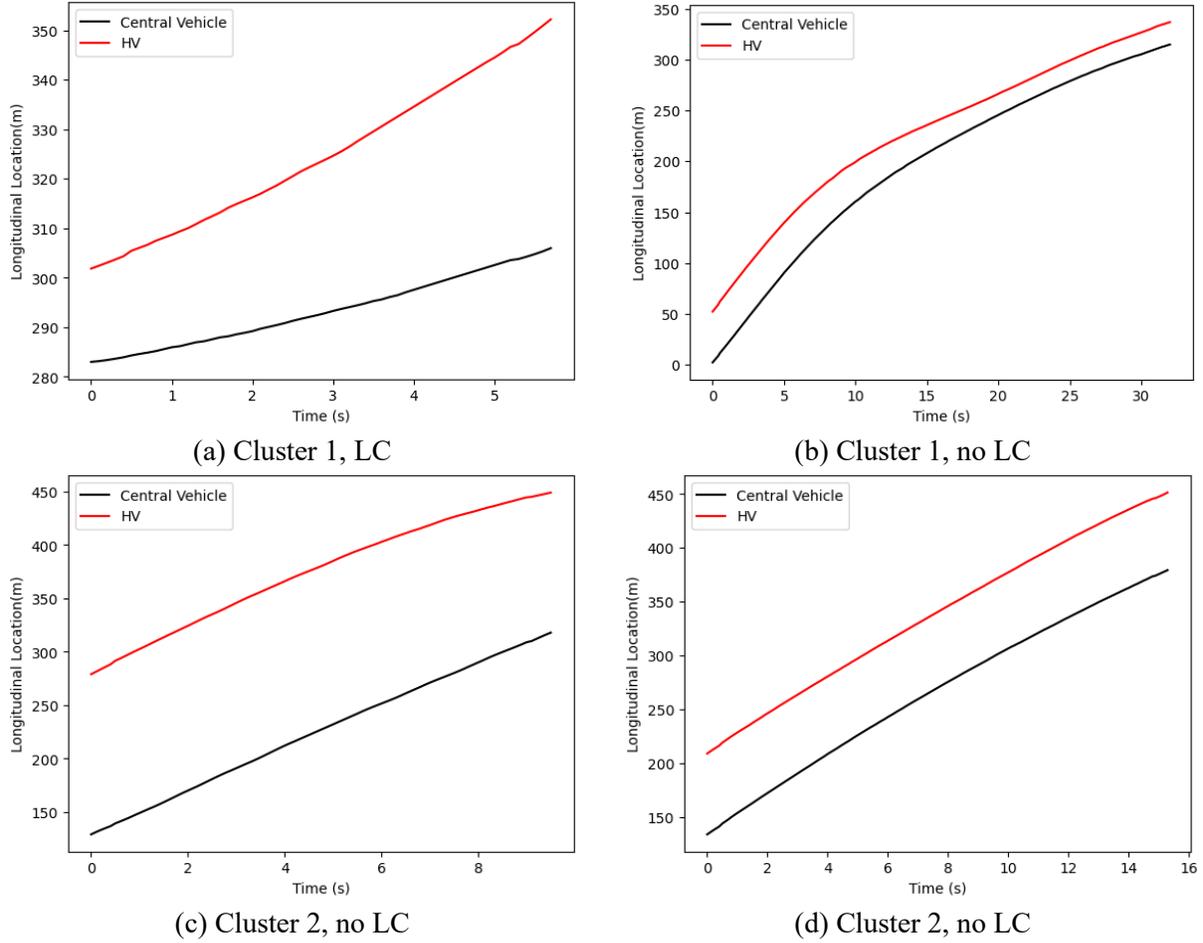

(a) Cluster 1, LC

(b) Cluster 1, no LC

(c) Cluster 2, no LC

(d) Cluster 2, no LC

**Figure 4** Example Trajectory Pairs

## DRIFT DIFFUSION MODEL

In our study, the DDM is used to model the decision-making process of vehicles that are behind HVs and have an intention to LC. Compared to existing models summarized in Rahman et al. (*16*), the DDM has the advantage of dynamically capturing the effects of traffic environment on the driver's LC decision-making process. In addition, a probability distribution of the time to perform a LC for each vehicle can be obtained from the DDM, which potentially makes LCs more forecastable. This section introduces the formulation of a generalized DDM, discusses the environmental variables used in DDM, and presents the log-likelihood function.

### General Formulation

Before elaborating on the model specification of the DDM used in our study, it is important to understand how DDM guides decision-making processes in general. The DDM formulates the process to eventually make one out of several decisions as evidence accumulation over time (*14; 26; 35*). In all, a DDM has four key components: the initial evidence, the drift rate, the decision-making threshold for each alternative over time, and noise. Eq.2 shows the rate of change of the evidence $A(t)$ over time, where $\mu(t)$ is the time-dependent drift rate that may depend on $A(t)$ and the environmental variables $X(t)$. $\zeta(t)$ is the noise term which is typically drawn from some random distribution and represents the diffusion process.

$$\frac{dA(t)}{dt} = \mu(t) + \zeta(t) \qquad \text{Eq.2}$$

Figure 5 shows an example DDM with one available decision whose evidence threshold is a constant function $A^+(t) = 1$ at a time discretion of every 0.01s. The drift rate $\mu(t)$ slowly decreases over time, and the noise $\zeta(t)$ is drawn from a normal distribution with mean 0 and standard deviation 0.25. The expected evidence $\mathrm{E}[A(t)]$ is calculated by accumulating the drift rates over time. Figure 5(c) shows the probability of reaching $A^+(t)$ over time, and highest probability is reached when $E[A(t)] = A^+(t)$.

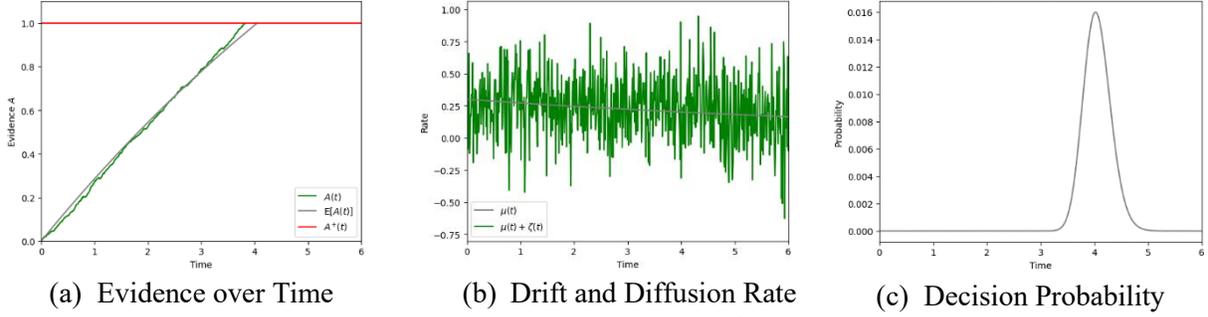

| (a) Evidence over Time | (b) Drift and Diffusion Rate | (c) Decision Probability |

**Figure 5** DDM Example

Now, we denote the decision-making time as $t_D$, which is the earliest time where the evidence threshold is reached, shown by Eq.3. The information at time later than $t_D$ is irrelevant to decision making.

$$t_D = \min \ \{ \ t \ | \ A(t) \geq A^+(t) \ \} \qquad \text{Eq.3}$$

For our LC model, we assume the DDM follows a Wiener Process where the drift rate is independent from the current evidence and the noise terms follows a time-independent normal distribution. This model structure is used because of its analytical tractability (26). Since the trajectory data is provided every 0.1s, Eq.4 is used to perform a discretization.

$$A(t_{i+1}) = A(t_i) + \mu(X(t_i))\Delta t + \sigma Z(t_i)\sqrt{\Delta t} \qquad \text{Eq.4}$$

$\Delta t$ is the size of each time step, which is 0.1s. $X(t_i)$ is the environmental variable that determines the drift rate at $t_i$ and will be discussed in "Model Specification". $Z(t_i)$ is a random scalar obtained by a standard normal distribution with mean 0 and standard variation 1. The diffusion process has a standard deviation of $\frac{\sigma}{\sqrt{\Delta t}}$. The diffusion term is expressed as $\sigma Z(t_i)\sqrt{\Delta t}$ to facilitate further derivations.

Now, we define $P^+(t_i)$ as the probability where threshold $A^+(t_i)$ is first reached at the time-step $t_i$. $P^+(t_i)$ can be approximated with the recursion shown by Eq.5 (26).

$$P^+(t_i) = -2\Psi(A^+(t_i), t_i \mid A(t_0), t_0) + 2\,\Delta t \sum_{k=1}^{i-1} P^+(t_k)\,\Psi(A^+(t_k), t_i \mid A(t_k), t_k\,) \qquad \text{Eq.5}$$

$\Psi$ is the kernel function defined by Eq.6.

$$\Psi(A^+(t), t \mid y, s) = \frac{f[A^+(t), t \mid y, s]}{2}\left(\frac{dA^+(t)}{dt} - \mu(t) - \frac{A^+(t) - y - \int_s^t \mu(X(r))dr}{t - s}\right) \qquad \text{Eq.6}$$

We furthermore define a state of the DD process as an ordered pair of time and evidence denoted by $(t, A(t))$. Then, $f[A^+(t), t \mid y, s]$, as defined by Eq.7, is the transition density, which is the likelihood of transition from the state $(s, y)$ to $(t, A^+(t))$.

$$f[A^+(t), t|y, s] = \frac{1}{\sqrt{2\pi\sigma^2(t-s)}} e^{-\frac{(A^+(t)-y-\int_s^t \mu(X(r))dr)^2}{2\sigma^2(t-s)}} \qquad \text{Eq.7}$$

Eq.8 approximates the integrals of the drift rates in Eq.6 and Eq.7 through discretization.

$$\int_{t_j}^{t_i} \mu(X(r))dr \sim \Delta t \sum_{k=j}^{i} \mu(X(t_k)) \qquad \text{Eq.8}$$

Eq.5, Eq.6 and Eq.7 can be derived from the Kolmogorov Equations and the trapezoidal approximation, and the detailed proof can be found in Richter et al. (*26*) and Smith (*35*). Finally, we assume that the initial condition of $P^+(t_0) = 0$ is satisfied for Eq.5, which means that the central vehicles do not perform a LC at the first time step of the trajectory pairs.

## Model Specification

Given the general DDM framework, we elaborate on the initial evidence for each central vehicle when it is first observed behind an HV and the environmental variables that determine the drift rate over time. We denote the initial evidence of a central vehicle $n$ as $A_n(t_0)$ and the environmental variables when vehicle $n$ is evaluating a LC to direction $d$ as $X_{n,d}(t_i)$.

It is important to realize the heterogeneity of the initial evidence $A_n(t_0)$ for each central vehicle because the urgency of performing a LC may differ. As discussed in "Sample Extraction" section, the start time of the HV-car trajectory pairs may not be when the car starts to be behind a HV. In fact, it is possible that earlier portions of the trajectory pairs are unobservable due to being out of the data collection site. It is expected that being closer to the HV may make a future LC more urgent, which can be reflected by higher initial evidence. To capture this phenomenon, we model the initial evidence as a decreasing function of the initial headway, as shown by Eq.9, where $h_n(t_0)$ is the initial headway between vehicle $n$ and the leading HV and $\alpha$ is a parameter to be obtained.

$$A_n(t_0) = 10 - \alpha h_n(t_0) \qquad \text{Eq.9}$$

For each central vehicle $n$, the evidence $A_n(t)$ accumulates over time with drift rates determined by the environmental variables. If the central vehicle has two adjacent lanes, LCs to both directions are considered and two independent DD processes take place in parallel. At time $t_i$ and given a possible LC direction $d$, we denote the evidence, the decision threshold, the environmental variables, and the drift rate as $A_{n,d}(t_i)$, $A_d^+(t_i)$, $X_{n,d}(t_i)$ and $\mu(X_{n,d}(t_i))$, respectively.

Eq.10 defines the 4 environmental variables used in this study. $G_{F,n,d}(t_i)$ is the adjacent follow gap as shown by Figure 2. $V_{n,d}(t_i)$ and $V_{HV,n}(t_i)$ are the speed of the adjacent leader and the HV respectively. Now we recall that $G_{L,n,d}(t_i)$ is the adjacent leader gap, then $\delta_{G,n,d}(t_i)$ is the dummy variable which equals 1 if the total gap size on the adjacent lane $G_{F,n,d}(t_i) + G_{L,n,d}(t_i)$ has increased compared to the last time step $t_{i-1}$ and 0 otherwise.

$$X_{n,d}(t_i) = [\, G_{F,n}(t_i), V_{n,d}(t_i), V_{HV,n}(t_i), \delta_{G,n,d}(t_i)] \qquad \text{Eq.10}$$

The drift rate is a function of the environmental variables and is shown by Eq.11, where $\beta_0, \beta_1, \beta_2, \beta_3$ are the parameters to be obtained. If $G_{F,n,d}(t_i) > G_{F,0}$, there is an incentive for the central vehicle to perform a LC. $V_{n,d}(t_i) - V_{HV,n}(t_i)$ is the speed difference between the adjacent leader and the leading HV, which approximates the current speed of the two lanes. A speed incentive exists if $V_{n,d}(t_i) > V_{HV,n}(t_i)$. The arctan activation function is applied to $G_{F,n,d}(t_i) - G_{F,0}$ and $V_{n,d}(t_i) - V_{HV,n}(t_i)$ to prevent obtaining large drift rates that can potentially introduce instability. It also reflects the diminishing marginal return of gap size and speed increment. Furthermore, the non-linearity introduced by arctan makes $\beta_0$ and $G_{F,0}$ estimable simultaneously, which can provide a meaningful interpretation. We expect $\beta_1, \beta_2$ and $\beta_3$ to be positive because a larger gap size, a higher speed differential between the adjacent lane and current lane, and an increasing gap size incentivize LCs. The reason why $G_{L,n,d}(t_i)$ is not included as an explicit input variable to this model is because its coefficient has been shown insignificant under a 95% confidence interval. This means that the adjacent lead gap is not a major consideration when the central vehicle is deciding when to LC, possibly because the central vehicle expects to be slowed down by the leading HV soon, which will eventually enlarge the adjacent lead gap size.

$$\mu(X_{n,d}(t_i)) = \beta_0 + \beta_1 \arctan(G_{F,n,d}(t_i) - G_{F,0}) + \beta_2 \arctan\left(V_{n,d}(t_i) - V_{HV,n}(t_i)\right) + \beta_3 \delta_{G,n,d}(t_i)$$ 

Eq.11

Finally, we assume a constant evidence threshold $A_d^+(t_i) = A_d^+ = 20$. According to Eq.9, the initial evidence is upper-bounded by 10 which is always less than $A_d^+$. In case when a central vehicle has two adjacent lanes, a LC is executed to the direction where the evidence threshold is exceeded earlier.

**Likelihood Function**

It is essential to customize a log-likelihood function for our sample because central vehicles may have 2 adjacent lanes and some of them fail to perform a LC. By maximizing the log-likelihood, we obtain the best DDM parameters.

As discussed in "Sample Extraction" section, a trajectory pair can end by four possible events. If it ends with the central vehicle performing a LC, we can reasonably conclude that a LC decision is made at the last time step. On the other hand, no LC decision is observed for the central vehicles with the other three ending events, and the evidence threshold is assumed never exceeded. Eq.12 shows the log-likelihood function for all the observations.

$$LL = \sum_n \left\{ \delta_n \log\left(P_{n,d_n}^+(t_{\max,n})\left(1 - F_{n,-d_n}(t_{\max,n})\right)\right) + (1 - \delta_n) \log\left(\left(1 - F_{n,-1}(t_{\max,n})\right)\left(1 - F_{n,1}(t_{\max,n})\right)\right) \right\}$$

Eq.12

$\delta_n$ is the dummy variable that equals 1 if the central vehicle executes a LC and 0 otherwise. $d_n$ the observed direction of LC for vehicle $n$ and equals -1 (1) if it is to the left (right). $t_{\max,n}$ is the last time step for the trajectory pair. $P_{n,d}^+(t_i)$ is the probability that the evidence threshold in direction $d$ is first passed at time step $t_i$, and $F_{n,d}^+(t_i)$ is the cumulative probability that the evidence threshold in direction $d$ has been passed before or at $t_i$. If a LC is observed for vehicle $n$, the likelihood of performing a LC at $t_{\max,n}$ is the joint probability where the evidence is reached exactly at $t_{\max,n}$ for direction $d_n$ but has never been reached for $-d_n$. With the independence assumption of the two DDM processes, we express the joint probability as $P_{n,d_n}^+(t_{\max,n})\left(1 - F_{n,-d_n}(t_{\max,n})\right)$. If no LC is made, then the evidence has never been reached for either direction, and this probability can be expressed as $\left(1 - F_{n,-1}(t_{\max,n})\right)\left(1 - F_{n,1}(t_{\max,n})\right)$. Finally, if a vehicle only has one adjacent lane, then all LC probability values associated with the direction where an adjacent lane does not exist are 0.

**RESULT**

This section discusses the fitting results of the DDM with central vehicles in cluster 1, which are those that likely have LC intention. To obtain the parameters jointly, we use the SciPy package in Python with a BFGS Solver (*34*). Table 1 shows the parameter values and the goodness of fit. In addition to parameters in Eq.9 and Eq.11, the value of $\sigma$, which is associated with the diffusion process, is also to be determined. We obtain one set of parameters for all the central vehicles in cluster 1.

Based on Table 1, we find that all coefficients except $\beta_0$, the constant term of the drift rate in Eq.11, are statistically significant under the 95% confidence level. The value of $\alpha$ indicates that when the initial headway between the central vehicle and the HV increases by 1s, the initial evidence decreases by 0.3267 on average. When the initial headway is higher, there is more room for the evidence to increase until the threshold is reached. This confirms our assumption that when the headway to the HV is smaller, drivers may get more impatient, and a LC becomes more urgent.

| Parameter | Mean | t-score | p-value |
|:---:|:---:|:---:|:---:|
| $\alpha$ | 0.3267 | 3.2310 | 0.0013 |
| $\beta_0$ | -0.2313 | -1.8126 | 0.0710 |
| $\beta_1$ | 0.1824 | 3.8720 | 0.0001 |
| $\beta_2$ | 0.0994 | 2.1017 | 0.0365 |
| $\beta_3$ | 0.7376 | 3.2500 | 0.0013 |
| $G_{F,0}$ | 16.7484 | 6.3591 | 0.0000 |
| $\sigma$ | 1.9147 | 10.6516 | 0.0000 |
| **Sample Size** | 268 | | |
| **Log-likelihood** | -309.5045 | | |

**Table 1** Model Parameters

The values of $\beta_1$ and $G_{F,0}$ indicate that the drift rate increases when the adjacent follow gap is greater than 16.7484m and decreases otherwise. This also implies that 16.7484m is the minimum adjacent follow gap that is desirable to perform a LC for an average central vehicle. If the adjacent follow gap size is smaller, then it tends to create a disincentive to perform a LC and lowers the drift rate. A positive $\beta_2$ suggests that a higher speed of the adjacent leader compared to the HV increases the drift rate and makes a LC sooner. A relatively high value of $\beta_3$ indicates that whether the gap is enlarging plays a crucial role in the central vehicle's decision making. If the total gap size on the adjacent lane increases compared to last time step, the drift rate is 0.7376 higher than otherwise. $\beta_0$ shows the average base drift rate and its negative value may indicate that when the adjacent gap size is not increasing and the incentives of gap and speed are both 0, the central vehicle is less likely to make a LC. This shows an adaptation process that if the condition to perform a LC is only marginally favorable, the driver of the central vehicle still gets more used to following the HV or may give up the LC eventually. The standard deviation of the diffusion process $\frac{\sigma}{\sqrt{\Delta t}}$ is 3.778, which is relatively large compared to the magnitude of the diffusion rate. This is possibly because most central vehicles failed to perform LCS and may also indicate high variability in the central vehicles' decision-making processes.

Now, we take a close look into the fitted DD process of three selected central vehicles by examining their drift rate and LC probability at each time-step, as shown by Figure 6, Figure 7 and Figure 8. To better visualize the trend, the drift rates and LC probabilities are smoothened by a Savitzky-Golay filter (*36*). We note that the sum of LC probabilities for each central vehicle is less than one because there is a likelihood of not performing a LC. Since the time step of 0.1s is low compared to the total duration, the LC probabilities are low at each time step. In addition, the relatively large proportion of central vehicles that have not performed a LC also makes the LC probability low in general.

According to Figure 6, the adjacent follow gap size is initially unsatisfying which leads to a negative drift rate. However, the adjacent follower decelerates and enlarges the gap, which makes the drift rate positive, leading to the central vehicle's LC at around 16s. In Figure 7 and Figure 8, the central vehicle has 2 adjacent lanes, which means that 2 independent DD processes are used in parallel. For both vehicles, the adjacent leaders and followers on the two adjacent lanes have changed frequently which makes the drift rate fluctuate. In Figure 7(c), local maximum probability is reached at 6s for a LC to the right lane and 7s to the left lane, even though no LC are made at these peaks. According to the trajectories in Figure 7(a), we find that at these two time steps, the adjacent follow gap is relatively large, which is a LC incentive in our DDM formulation. However, the adjacent leaders at these two probability maxima are close, and a LC is not safe to be performed. This observation indicates a limitation of our model, where LC executions may be predicted even when they are not feasible. At around 16s, however, the follow gap on the right lane is large enough and increasing, yet the left lane has a slower leader. This facilitates executing a LC to the right, which is also temporally close to the global maximum of the LC probability to the right lane. In Figure 8, no LC is eventually made. The drift rate to the left tends to be negative, because the left lane is experiencing LCs from the on ramp and the high local density discourages more LCs into it. On the other hand, the LC probability of the right lane drastically increases after 25 seconds because the right follow gap is enlarging, and the right leader is faster. This may imply that a LC to the right may likely be made slightly downstream soon, even though it is unobserved.

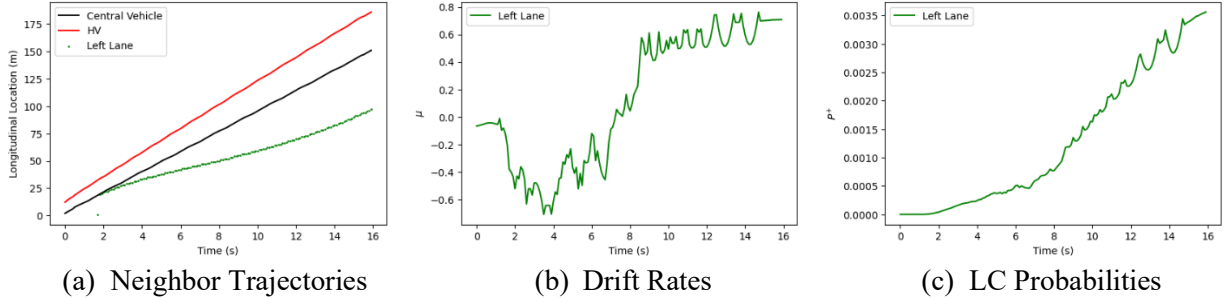

(a) Neighbor Trajectories       (b) Drift Rates       (c) LC Probabilities

**Figure 6** DD Process Example: 1 Adjacent Lane, LC Made to the Left

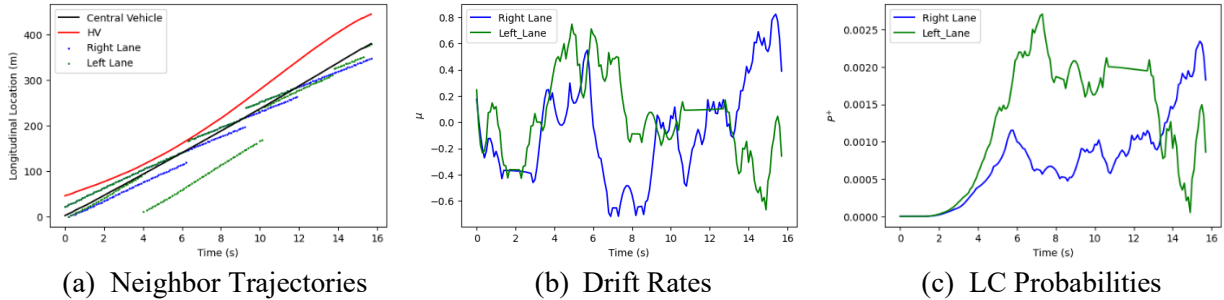

(a) Neighbor Trajectories       (b) Drift Rates       (c) LC Probabilities

**Figure 7** DD Process Example: 2 Adjacent Lanes, LC Made to the Right

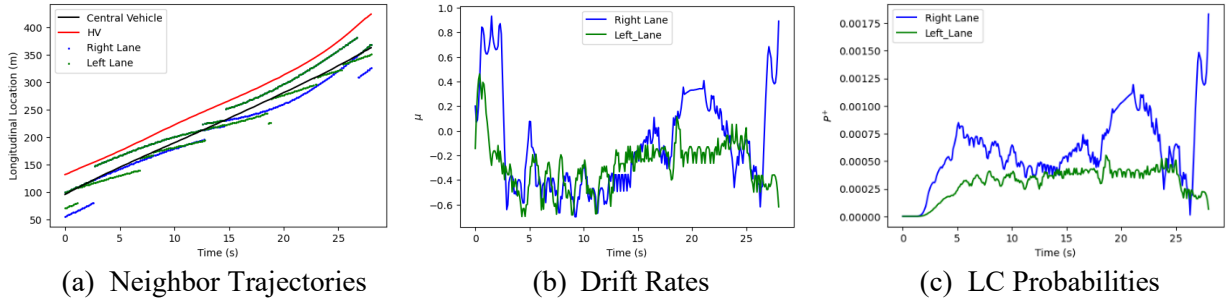

(a) Neighbor Trajectories       (b) Drift Rates       (c) LC Probabilities

**Figure 8** DD Process Example: 2 Adjacent Lanes, LC not Made

## CONCLUSION AND FUTURE RESEARCH

In this study, we use a DDM model to analyze the LC decision-making process of cars behind HVs. Our DDM assumes a Wiener process where the drift rate only depends on the central vehicles' surrounding traffic environment and the initial evidence only depends on the initial headway. If a central vehicle has two adjacent lanes, two independent DDMs take place in parallel, and a LC is executed to the direction where the evidence of the DDM first reaches the threshold. To obtain vehicles with LC intention behind HVs, we extracted HV-car trajectory pairs in the TGSIM I-395 dataset and used K-means to group them into two clusters based on longitudinal features. We find that all central vehicles that have performed a LC fall in the same cluster where the central vehicles behave more differently to the leading HV, and the lead gap is smaller. This implies that central vehicles in this cluster likely have LC intention. In all, the DDM is fitted with 265 central vehicles with LC intention, among which 41 have performed a LC. The fitted parameters indicate that an increasing total gap size on the adjacent lane over time, a larger adjacent follow gap size, and a higher speed difference between the adjacent leader and the HV contribute to a higher drift rate toward a LC. In addition, a smaller initial headway to the HV is associated with higher initial evidence, which indicates more impatience and may lead to a LC sooner.

While using DDM is a novel approach to model LCs and our model yields interpretable parameters, our study has a few limitations. First, the clustering process is based on the longitudinal driving behavior and does not ensure that the included central vehicles all have LC intention. To solve this problem, we would treat the LC intention as a latent variable and jointly perform clustering and DDM fitting. Second, the only heterogeneity considered in the DDM is the initial evidence of the central vehicle. However, drivers are in fact heterogeneous and may evaluate the same surrounding traffic state differently. For example, aggressive drivers are more willing to merge into gaps that are considered too small by cautious drivers. This may also explain the relatively large diffusion term. In the future, we would introduce random coefficients across drivers and time-dependent diffusion into the DDM. Furthermore, our DDM does not guarantee that the predicted LC execution time is feasible. To eliminate this problem, we would incorporate safety constraints into the decision-making threshold formulation, so that it cannot be reached when a LC is not safe. Finally, the DDM does not consider longitudinal motion of the central vehicles. However, a vehicle may decide to adjust their speed to align with a gap. To address this limitation, we will apply the DDM and a longitudinal model jointly to provide a wholistic framework for LCs of vehicles behind HVs.

## ACKNOWLEDGEMENTS
The authors of this paper sincerely thank Alireza Talebpour and Ying Chen for their kind assistance and support.

## CONTRIBUTIONS
The authors claim the contribution to this paper as the following: study conception and modelling: Nachuan Li, Hani Mahmassani, Sue Ahn, Anupam Srivastava; data processing and interpretation of results: Nachuan Li, Hani Mahmassani; manuscript preparation: Nachuan Li, Hani Mahmassani, Sue Ahn, Anupam Srivastava.